%% file: main.tex
\documentclass[twocolumn]{aastex62}

\input{commands}

\usepackage{amsmath}

\usepackage{lineno}

\graphicspath{{./}{figures/}}

\accepted{February 9, 2019}
\submitjournal{ApJL}

\shorttitle{Search for Optical Emission from GW170814 with DECam}
\shortauthors{Z.~Doctor, R.~Kessler, et al.}

\begin{document}

\title{A Search for Optical Emission from Binary-Black-Hole Merger GW170814 with the Dark Energy Camera}

\input{DES-2018-0335_author_list}

\begin{abstract}

\input{abstract}

\end{abstract}


\newpage
\section{Introduction} \label{sec:intro}
\input{intro}

\section{Search and Light Curves} \label{sec:search}
\input{search}


\section{Analysis} \label{sec:analysis}
\input{analysis}

\section{Results} \label{sec:results}
\input{results}

\section{Discussion} \label{sec:discussion}
\input{discussion}

\section{Conclusion} \label{sec:conclusion}
\input{conclusion}

\section{Acknowledgments} 
\input{ack}

\bibliography{main.bib}



\end{document}

%% file: commands.tex
\newcommand{\GW}{{\small GW}}
\newcommand{\GWs}{{\small GWs}}
\newcommand{\BBH}{{\small BBH}}
\newcommand{\BBHs}{{\small BBHs}}
\newcommand{\LIGO}{{\small LIGO}}
\newcommand{\EM}{{\small EM}}
\newcommand{\DECam}{{\small DEC}am}
\newcommand{\Bayestar}{\texttt{Bayestar}}
\newcommand{\DES}{{\small DES}}
\newcommand{\diffimg}{\texttt{DiffImg}}
\newcommand{\SN}{{\small SN}}

\newcommand{\snana}{\texttt{SNANA}}

\newcommand{\IDblind}{2}
\newcommand{\IDcontrol}{1}
\newcommand{\area}{225}
\newcommand{\LVC}{{\small LVC}}


%% file: DES-2018-0335_author_list.tex

\author{Z.~Doctor}
\affiliation{Department of Physics, University of Chicago, Chicago, IL 60637, USA}
\affiliation{Kavli Institute for Cosmological Physics, University of Chicago, Chicago, IL 60637, USA}
\author{R.~Kessler}
\affiliation{Kavli Institute for Cosmological Physics, University of Chicago, Chicago, IL 60637, USA}
\affiliation{Department of Astronomy and Astrophysics, University of Chicago, Chicago, IL 60637, USA}
\author{K.~Herner}
\affiliation{Fermi National Accelerator Laboratory, P. O. Box 500, Batavia, IL 60510, USA}
\author{A.~Palmese}
\affiliation{Fermi National Accelerator Laboratory, P. O. Box 500, Batavia, IL 60510, USA}
\author{M.~Soares-Santos}
\affiliation{Brandeis University, Physics Department, 415 South Street, Waltham MA 02453}
\author{J.~Annis}
\affiliation{Fermi National Accelerator Laboratory, P. O. Box 500, Batavia, IL 60510, USA}
\author{D.~Brout}
\affiliation{Department of Physics and Astronomy, University of Pennsylvania, Philadelphia, PA 19104, USA}
\author{D.~E.~Holz}
\affiliation{Department of Physics, University of Chicago, Chicago, IL 60637, USA}
\affiliation{Kavli Institute for Cosmological Physics, University of Chicago, Chicago, IL 60637, USA}
\affiliation{Department of Astronomy and Astrophysics, University of Chicago, Chicago, IL 60637, USA}
\affiliation{Enrico Fermi Institute, University of Chicago, Chicago, IL 60637, USA}
\author{M.~Sako}
\affiliation{Department of Physics and Astronomy, University of Pennsylvania, Philadelphia, PA 19104, USA}
\author{A.~Rest}
\affiliation{Department of Physics and Astronomy, The Johns Hopkins University, 3400 North Charles Street, Baltimore, MD 21218, USA}
\affiliation{Space Telescope Science Institute, 3700 San Martin Drive, Baltimore, MD 21218, USA}
\author{P.~Cowperthwaite}
\affiliation{Carnegie Observatories, 813 Santa Barbara St., Pasadena, CA 91104}
\author{E.~Berger}
\affiliation{Harvard-Smithsonian Center for Astrophysics, 60 Garden Street, Cambridge, MA 02138, USA}
\author{R.~J.~Foley}
\affiliation{Santa Cruz Institute for Particle Physics, Santa Cruz, CA 95064, USA}
\author{C.~J.~Conselice}
\affiliation{Centre for Astronomy and Particle Theory, School of Physics \& Astronomy, University of Nottingham, Nottingham, NG7 2RD UK}
\author{M.S.S.~Gill}
\affiliation{SLAC National Accelerator Laboratory, Menlo Park, CA 94025, USA}
\affiliation{Kavli Institute for Particle Astrophysics \& Cosmology, P. O. Box 2450, Stanford University, Stanford, CA 94305, USA}
\author{S.~Allam}
\affiliation{Fermi National Accelerator Laboratory, P. O. Box 500, Batavia, IL 60510, USA}
\author{E.~Balbinot}
\affiliation{Kapteyn Astronomical Institute, University of Groningen, Landleven 12, 9747 AD Groningen, The Netherlands}
\author{R.~E.~Butler}
\affiliation{Department of Astronomy, Indiana University, 727 E. Third Street, Bloomington, IN 47405, USA}
\author{H.-Y.~Chen}
\affiliation{Black Hole Initiative, Harvard University, Cambridge, MA 02138, USA}
\author{R.~Chornock}
\affiliation{Astrophysical Institute, Department of Physics and Astronomy, 251B Clippinger Lab, Ohio University, Athens, OH 45701, USA}
\author{E.~Cook}
\affiliation{George P. and Cynthia Woods Mitchell Institute for Fundamental Physics and Astronomy, and Department of Physics and Astronomy, Texas A\&M University, College Station, TX 77843,  USA}
\author{H.~T.~Diehl}
\affiliation{Fermi National Accelerator Laboratory, P. O. Box 500, Batavia, IL 60510, USA}
\author{B.~Farr}
\affiliation{Department of Physics, University of Oregon, Eugene, OR 97403, USA}
\author{W.~Fong}
\affiliation{Center for Interdisciplinary Exploration and Research in Astrophysics (CIERA) and Department of Physics and Astronomy, Northwestern University, Evanston, IL 60208}
\author{J.~Frieman}
\affiliation{Fermi National Accelerator Laboratory, P. O. Box 500, Batavia, IL 60510, USA}
\affiliation{Kavli Institute for Cosmological Physics, University of Chicago, Chicago, IL 60637, USA}
\author{C.~Fryer}
\affiliation{Center for Theoretical Astrophysics, Los Alamos National Laboratory, Los Alamos, NM 87545, USA}
\author{J.~Garc\'ia-Bellido}
\affiliation{Instituto de Fisica Teorica UAM/CSIC, Universidad Autonoma de Madrid, 28049 Madrid, Spain}
\author{R.~Margutti}
\affiliation{Center for Interdisciplinary Exploration and Research in Astrophysics (CIERA) and Department of Physics and Astronomy, Northwestern University, Evanston, IL 60208}
\author{J.~L.~Marshall}
\affiliation{George P. and Cynthia Woods Mitchell Institute for Fundamental Physics and Astronomy, and Department of Physics and Astronomy, Texas A\&M University, College Station, TX 77843,  USA}
\author{T.~Matheson}
\affiliation{National Optical Astronomy Observatory, 950 North Cherry Avenue, Tucson, AZ 85719}
\author{B.~D.~Metzger}
\affiliation{Department of Physics and Columbia Astrophysics Laboratory, Columbia University, New York, NY 10027, USA}
\author{M.~Nicholl}
\affiliation{Institute for Astronomy, University of Edinburgh, Royal Observatory, Blackford Hill, Edinburgh EH9 3HJ, UK}
\author{F.~Paz-Chinch\'on}
\affiliation{National Center for Supercomputing Applications, University of Illinois at Urbana-Champaign, 1205 W Clark St. Urbana, IL. 61801}
\author{S.~Salim}
\affiliation{Department of Astronomy, Indiana University, 727 E. Third Street, Bloomington, IN 47405, USA}
\author{M.~Sauseda}
\affiliation{Department of Physics and Astronomy, Texas A\&M University, College Station, TX 77843,  USA}
\author{L.~F.~Secco}
\affiliation{Department of Physics and Astronomy, University of Pennsylvania, Philadelphia, PA 19104, USA}
\author{R.~C.~Smith}
\affiliation{Cerro Tololo Inter-American Observatory, National Optical Astronomy Observatory, Casilla 603, La Serena, Chile}
\author{N.~Smith}
\affiliation{Steward Observatory, University of Arizona, 933 N. Cherry Ave., Tucson, AZ 85721, USA}
\author{A.~K.~Vivas}
\affiliation{Cerro Tololo Inter-American Observatory, National Optical Astronomy Observatory, Casilla 603, La Serena, Chile}
\author{D.~L.~Tucker}
\affiliation{Fermi National Accelerator Laboratory, P. O. Box 500, Batavia, IL 60510, USA}
\author{T.~M.~C.~Abbott}
\affiliation{Cerro Tololo Inter-American Observatory, National Optical Astronomy Observatory, Casilla 603, La Serena, Chile}
\author{S.~Avila}
\affiliation{Institute of Cosmology and Gravitation, University of Portsmouth, Portsmouth, PO1 3FX, UK}
\author{K.~Bechtol}
\affiliation{LSST, 933 North Cherry Avenue, Tucson, AZ 85721, USA}
\affiliation{Physics Department, 2320 Chamberlin Hall, University of Wisconsin-Madison, 1150 University Avenue Madison, WI  53706-1390}
\author{E.~Bertin}
\affiliation{CNRS, UMR 7095, Institut d'Astrophysique de Paris, F-75014, Paris, France}
\affiliation{Sorbonne Universit\'es, UPMC Univ Paris 06, UMR 7095, Institut d'Astrophysique de Paris, F-75014, Paris, France}
\author{D.~Brooks}
\affiliation{Department of Physics \& Astronomy, University College London, Gower Street, London, WC1E 6BT, UK}
\author{E.~Buckley-Geer}
\affiliation{Fermi National Accelerator Laboratory, P. O. Box 500, Batavia, IL 60510, USA}
\author{D.~L.~Burke}
\affiliation{Kavli Institute for Particle Astrophysics \& Cosmology, P. O. Box 2450, Stanford University, Stanford, CA 94305, USA}
\affiliation{SLAC National Accelerator Laboratory, Menlo Park, CA 94025, USA}
\author{A.~Carnero~Rosell}
\affiliation{Centro de Investigaciones Energ\'eticas, Medioambientales y Tecnol\'ogicas (CIEMAT), Madrid, Spain}
\affiliation{Laborat\'orio Interinstitucional de e-Astronomia - LIneA, Rua Gal. Jos\'e Cristino 77, Rio de Janeiro, RJ - 20921-400, Brazil}
\author{M.~Carrasco~Kind}
\affiliation{Department of Astronomy, University of Illinois at Urbana-Champaign, 1002 W. Green Street, Urbana, IL 61801, USA}
\affiliation{National Center for Supercomputing Applications, 1205 West Clark St., Urbana, IL 61801, USA}
\author{J.~Carretero}
\affiliation{Institut de F\'{\i}sica d'Altes Energies (IFAE), The Barcelona Institute of Science and Technology, Campus UAB, 08193 Bellaterra (Barcelona) Spain}
\author{F.~J.~Castander}
\affiliation{Institut d'Estudis Espacials de Catalunya (IEEC), 08034 Barcelona, Spain}
\affiliation{Institute of Space Sciences (ICE, CSIC),  Campus UAB, Carrer de Can Magrans, s/n,  08193 Barcelona, Spain}
\author{C.~B.~D'Andrea}
\affiliation{Department of Physics and Astronomy, University of Pennsylvania, Philadelphia, PA 19104, USA}
\author{L.~N.~da Costa}
\affiliation{Laborat\'orio Interinstitucional de e-Astronomia - LIneA, Rua Gal. Jos\'e Cristino 77, Rio de Janeiro, RJ - 20921-400, Brazil}
\affiliation{Observat\'orio Nacional, Rua Gal. Jos\'e Cristino 77, Rio de Janeiro, RJ - 20921-400, Brazil}
\author{J.~De~Vicente}
\affiliation{Centro de Investigaciones Energ\'eticas, Medioambientales y Tecnol\'ogicas (CIEMAT), Madrid, Spain}
\author{S.~Desai}
\affiliation{Department of Physics, IIT Hyderabad, Kandi, Telangana 502285, India}
\author{P.~Doel}
\affiliation{Department of Physics \& Astronomy, University College London, Gower Street, London, WC1E 6BT, UK}
\author{B.~Flaugher}
\affiliation{Fermi National Accelerator Laboratory, P. O. Box 500, Batavia, IL 60510, USA}
\author{P.~Fosalba}
\affiliation{Institut d'Estudis Espacials de Catalunya (IEEC), 08034 Barcelona, Spain}
\affiliation{Institute of Space Sciences (ICE, CSIC),  Campus UAB, Carrer de Can Magrans, s/n,  08193 Barcelona, Spain}
\author{E.~Gaztanaga}
\affiliation{Institut d'Estudis Espacials de Catalunya (IEEC), 08034 Barcelona, Spain}
\affiliation{Institute of Space Sciences (ICE, CSIC),  Campus UAB, Carrer de Can Magrans, s/n,  08193 Barcelona, Spain}
\author{D.~W.~Gerdes}
\affiliation{Department of Astronomy, University of Michigan, Ann Arbor, MI 48109, USA}
\affiliation{Department of Physics, University of Michigan, Ann Arbor, MI 48109, USA}
\author{D.~A.~Goldstein}
\affiliation{California Institute of Technology, 1200 East California Blvd, MC 249-17, Pasadena, CA 91125, USA}
\author{D.~Gruen}
\affiliation{Kavli Institute for Particle Astrophysics \& Cosmology, P. O. Box 2450, Stanford University, Stanford, CA 94305, USA}
\affiliation{SLAC National Accelerator Laboratory, Menlo Park, CA 94025, USA}
\author{R.~A.~Gruendl}
\affiliation{Department of Astronomy, University of Illinois at Urbana-Champaign, 1002 W. Green Street, Urbana, IL 61801, USA}
\affiliation{National Center for Supercomputing Applications, 1205 West Clark St., Urbana, IL 61801, USA}
\author{G.~Gutierrez}
\affiliation{Fermi National Accelerator Laboratory, P. O. Box 500, Batavia, IL 60510, USA}
\author{W.~G.~Hartley}
\affiliation{Department of Physics \& Astronomy, University College London, Gower Street, London, WC1E 6BT, UK}
\affiliation{Department of Physics, ETH Zurich, Wolfgang-Pauli-Strasse 16, CH-8093 Zurich, Switzerland}
\author{D.~L.~Hollowood}
\affiliation{Santa Cruz Institute for Particle Physics, Santa Cruz, CA 95064, USA}
\author{K.~Honscheid}
\affiliation{Center for Cosmology and Astro-Particle Physics, The Ohio State University, Columbus, OH 43210, USA}
\affiliation{Department of Physics, The Ohio State University, Columbus, OH 43210, USA}
\author{B.~Hoyle}
\affiliation{Max Planck Institute for Extraterrestrial Physics, Giessenbachstrasse, 85748 Garching, Germany}
\affiliation{Universit\"ats-Sternwarte, Fakult\"at f\"ur Physik, Ludwig-Maximilians Universit\"at M\"unchen, Scheinerstr. 1, 81679 M\"unchen, Germany}
\author{D.~J.~James}
\affiliation{Harvard-Smithsonian Center for Astrophysics, Cambridge, MA 02138, USA}
\author{T.~Jeltema}
\affiliation{Santa Cruz Institute for Particle Physics, Santa Cruz, CA 95064, USA}
\author{S.~Kent}
\affiliation{Fermi National Accelerator Laboratory, P. O. Box 500, Batavia, IL 60510, USA}
\affiliation{Kavli Institute for Cosmological Physics, University of Chicago, Chicago, IL 60637, USA}
\author{K.~Kuehn}
\affiliation{Australian Astronomical Optics, Macquarie University, North Ryde, NSW 2113, Australia}
\author{N.~Kuropatkin}
\affiliation{Fermi National Accelerator Laboratory, P. O. Box 500, Batavia, IL 60510, USA}
\author{O.~Lahav}
\affiliation{Department of Physics \& Astronomy, University College London, Gower Street, London, WC1E 6BT, UK}
\author{M.~Lima}
\affiliation{Departamento de F\'isica Matem\'atica, Instituto de F\'isica, Universidade de S\~ao Paulo, CP 66318, S\~ao Paulo, SP, 05314-970, Brazil}
\affiliation{Laborat\'orio Interinstitucional de e-Astronomia - LIneA, Rua Gal. Jos\'e Cristino 77, Rio de Janeiro, RJ - 20921-400, Brazil}
\author{M.~A.~G.~Maia}
\affiliation{Laborat\'orio Interinstitucional de e-Astronomia - LIneA, Rua Gal. Jos\'e Cristino 77, Rio de Janeiro, RJ - 20921-400, Brazil}
\affiliation{Observat\'orio Nacional, Rua Gal. Jos\'e Cristino 77, Rio de Janeiro, RJ - 20921-400, Brazil}
\author{M.~March}
\affiliation{Department of Physics and Astronomy, University of Pennsylvania, Philadelphia, PA 19104, USA}
\author{F.~Menanteau}
\affiliation{Department of Astronomy, University of Illinois at Urbana-Champaign, 1002 W. Green Street, Urbana, IL 61801, USA}
\affiliation{National Center for Supercomputing Applications, 1205 West Clark St., Urbana, IL 61801, USA}
\author{C.~J.~Miller}
\affiliation{Department of Astronomy, University of Michigan, Ann Arbor, MI 48109, USA}
\affiliation{Department of Physics, University of Michigan, Ann Arbor, MI 48109, USA}
\author{R.~Miquel}
\affiliation{Instituci\'o Catalana de Recerca i Estudis Avan\c{c}ats, E-08010 Barcelona, Spain}
\affiliation{Institut de F\'{\i}sica d'Altes Energies (IFAE), The Barcelona Institute of Science and Technology, Campus UAB, 08193 Bellaterra (Barcelona) Spain}
\author{E.~Neilsen}
\affiliation{Fermi National Accelerator Laboratory, P. O. Box 500, Batavia, IL 60510, USA}
\author{B.~Nord}
\affiliation{Fermi National Accelerator Laboratory, P. O. Box 500, Batavia, IL 60510, USA}
\author{R.~L.~C.~Ogando}
\affiliation{Laborat\'orio Interinstitucional de e-Astronomia - LIneA, Rua Gal. Jos\'e Cristino 77, Rio de Janeiro, RJ - 20921-400, Brazil}
\affiliation{Observat\'orio Nacional, Rua Gal. Jos\'e Cristino 77, Rio de Janeiro, RJ - 20921-400, Brazil}
\author{A.~A.~Plazas}
\affiliation{Jet Propulsion Laboratory, California Institute of Technology, 4800 Oak Grove Dr., Pasadena, CA 91109, USA}
\author{A.~Roodman}
\affiliation{Kavli Institute for Particle Astrophysics \& Cosmology, P. O. Box 2450, Stanford University, Stanford, CA 94305, USA}
\affiliation{SLAC National Accelerator Laboratory, Menlo Park, CA 94025, USA}
\author{E.~Sanchez}
\affiliation{Centro de Investigaciones Energ\'eticas, Medioambientales y Tecnol\'ogicas (CIEMAT), Madrid, Spain}
\author{V.~Scarpine}
\affiliation{Fermi National Accelerator Laboratory, P. O. Box 500, Batavia, IL 60510, USA}
\author{R.~Schindler}
\affiliation{SLAC National Accelerator Laboratory, Menlo Park, CA 94025, USA}
\author{M.~Schubnell}
\affiliation{Department of Physics, University of Michigan, Ann Arbor, MI 48109, USA}
\author{S.~Serrano}
\affiliation{Institut d'Estudis Espacials de Catalunya (IEEC), 08034 Barcelona, Spain}
\affiliation{Institute of Space Sciences (ICE, CSIC),  Campus UAB, Carrer de Can Magrans, s/n,  08193 Barcelona, Spain}
\author{I.~Sevilla-Noarbe}
\affiliation{Centro de Investigaciones Energ\'eticas, Medioambientales y Tecnol\'ogicas (CIEMAT), Madrid, Spain}
\author{M.~Smith}
\affiliation{School of Physics and Astronomy, University of Southampton,  Southampton, SO17 1BJ, UK}
\author{F.~Sobreira}
\affiliation{Instituto de F\'isica Gleb Wataghin, Universidade Estadual de Campinas, 13083-859, Campinas, SP, Brazil}
\affiliation{Laborat\'orio Interinstitucional de e-Astronomia - LIneA, Rua Gal. Jos\'e Cristino 77, Rio de Janeiro, RJ - 20921-400, Brazil}
\author{E.~Suchyta}
\affiliation{Computer Science and Mathematics Division, Oak Ridge National Laboratory, Oak Ridge, TN 37831}
\author{M.~E.~C.~Swanson}
\affiliation{National Center for Supercomputing Applications, 1205 West Clark St., Urbana, IL 61801, USA}
\author{G.~Tarle}
\affiliation{Department of Physics, University of Michigan, Ann Arbor, MI 48109, USA}
\author{D.~Thomas}
\affiliation{Institute of Cosmology and Gravitation, University of Portsmouth, Portsmouth, PO1 3FX, UK}
\author{A.~R.~Walker}
\affiliation{Cerro Tololo Inter-American Observatory, National Optical Astronomy Observatory, Casilla 603, La Serena, Chile}
\author{W.~Wester}
\affiliation{Fermi National Accelerator Laboratory, P. O. Box 500, Batavia, IL 60510, USA}

\collaboration{(DES Collaboration)}

%% file: abstract.tex
Binary black hole (\BBH) mergers found by the \LIGO{} and Virgo detectors are of immense scientific interest to the astrophysics community, but are considered unlikely to be sources of electromagnetic emission.  To test whether they have rapidly fading optical counterparts, we used the Dark Energy Camera to perform an $i$-band search for the \BBH{} merger \GW170814, the first gravitational wave detected by three interferometers.  The 87-deg$^2$ localization region (at 90\% confidence) centered in the Dark Energy Survey (\DES{}) footprint enabled us to image 86\% of the probable sky area to a depth of $i\sim 23$ mag and provide the most comprehensive dataset to search for \EM{} emission from \BBH{} mergers.  To identify candidates, we perform difference imaging with our search images and with templates from pre-existing \DES{} images. The analysis strategy and selection requirements were designed to remove supernovae and to identify transients that decline in the first two epochs. We find two candidates, each of which is spatially coincident with a star or a high-redshift galaxy in the \DES{} catalogs, and they are thus unlikely to be associated with \GW170814. Our search finds no candidates associated with \GW170814, disfavoring rapidly declining optical emission from \BBH{} mergers brighter than $i\sim 23$ mag ($L_{\rm optical} \sim 5\times10^{41}$ erg/s) 1-2 days after coalescence. In terms of GW sky map coverage, this is the most complete search for optical counterparts to \BBH{} mergers to date.

%% file: intro.tex
Since the first binary black hole (\BBH) merger detection in September, 2015 \citep{GW150914}, mergers of two black holes have become a mainstay of gravitational-wave (\GW) astrophysics.  The first five observed \BBHs{}, found only by the Hanford and Livingston Laser Interferometer Gravitational-Wave Observatory (\LIGO) detectors, offered significant astrophysical insight into the \BBH{} mass distribution and event rates \citep{GW150914,150914astro,GW151226,O1LIGO,GW170104,GW170608}.  For electromagnetic (EM) follow-up, however, the two LIGO detectors alone place poor constraints on the sky position, typically a few hundred deg$^2$.

To date, no compelling optical counterparts to \BBH{} mergers have been identified.  However, \citet{Stalder} found optical candidate ATLAS17aeu in their follow-up of \GW170104 and hypothesize a chance coincidence. Additionally, a weak gamma-ray burst in coincidence with GW150914 was reported in \cite{Connaughton2016}, but its association with GW150914 is still under dispute.  There are three (not mutually exclusive) reasons for non-detections: (1) the probable sky regions of previous \BBH{} detections were not searched comprehensively, (2) the \BBH{} emission could not be identified or distinguished from background transients, and/or (3) optical emission from \BBH{} mergers is non-existent or below the detectable threshold at the times of the existing observations.  Theoretical models have been proposed which could produce \EM{} signals \citep[e.g.][]{deMink, Loeb, Perna, Stone, McKernan}, but these models are highly speculative. With little theoretical guidance, there is a need for more complete searches for \BBH{} \EM{} emission while also controlling false-positive event rates.  Detection of \BBH{} \EM{} counterparts would be of immense scientific value \citep[e.g.][]{Phinney}, as it could constrain the formation environments of \BBHs{}, the behavior of matter in strong field gravity, and cosmological parameters such as the Hubble constant\footnote{Even without an optical counterpart to a \BBH, it is possible to measure the Hubble constant with a \BBH{} \GW{} sky map and galaxy catalog as in e.g.~\citet{Schutz,ChenH0,SSH0}.}.

Thus far, a number of optical follow-up campaigns have been conducted to search for \BBH{} counterparts \citep[e.g.][]{GPSmith,Stalder,Smartt151226,Cowperthwaite2016,SS2016,Yoshida,Morokuma,Smartt150914,Lipunov150914}. However, the large probable sky areas of the ``double-coincident'' LIGO detections (Hanford and Livingston detectors only) curtailed searches for \EM{} counterparts from \BBH{} mergers. For example, \citet{SS2016} observed 102 deg$^2$ of the \GW150914 high-probability sky region with the optical imager Dark Energy Camera, \citep[\DECam:][]{Flaugher}, corresponding to 38\% of the initial \LIGO{} sky map probability.  After accounting for the lack of existing images (templates) for difference-imaging, a shift in the sky map in a reanalysis of \LIGO{} data, and other efficiency losses, only 3\% of the probable \GW150914 sky area was searched and analyzed. Similarly, the \DECam{} follow-up campaign of \GW151226 reported in \citet{Cowperthwaite2016} covered 29 deg$^2$, just $\sim 2\%$ of the final \GW151226 high probability region.  In contrast, with the three-detector network including the Virgo interferometer, the smaller 28-deg$^2$ 90\% localization region of neutron-star merger \GW170817 enabled 81\% \DECam{} coverage of the final \LIGO-Virgo sky map and identification of the \EM{} counterpart \citep{GW170817,SS17}.  These searches were all performed in the $i$ and $z$ bands, requiring two tilings of the search area.  We note that these \DECam{} searches attempted to tile maximal sky map probability, but for the nearby events such as \GW170817, targeting based on galaxy catalogs can be successful \citep[e.g.][]{Coulter,Valenti,Arcavi}.

On August 14, 2017, the LIGO-Virgo Collaboration (\LVC), with the addition of the Virgo detector, made the first ``triple-coincident'' detection of \GWs{} from a \BBH{} event, \GW170814, and provided a much tighter constraint on the sky position of the source than those of previous \BBH{} detections \citep{GW170814,GWCatalog}. The detection of \GW170814, with its $87$-deg$^2$ 90\%-localization region, enabled our team to perform a comprehensive search of the sky area for \BBH{} merger optical counterparts and significantly improve our sensitivity to \BBH{} merger \EM{} emission models.

We report on our search for optical counterparts to \GW170814 using the Dark Energy Camera.  In \S\ref{sec:search},  we describe the parameters and cadence of our follow-up observations, which extended to 12 days after the \GW170814 trigger and covered 225 deg$^2$.  \S\ref{sec:analysis} describes the analysis.  Finally, \S\ref{sec:results} presents the results of the analysis, which we then comment on in \S\ref{sec:discussion} and \S\ref{sec:conclusion}.

%% file: search.tex
On August 14, 2017 at 10:30:43 UTC, the \LVC{} reported a signal consistent with the inspiral and merger of two black holes of masses $30.5_{-3.0}^{+5.7} M_\odot$ and $ 25.3_{-4.2}^{+2.8} M_\odot$ at a luminosity distance of $540_{-210}^{+130}$ Mpc and redshift\footnote{Assuming cosmology of \cite{Planck15}} $z= 0.12_{-.04}^{+.03}$ \citep{GW170814}.  \LIGO{} and Virgo sent out a \Bayestar{} sky map 2 hours after the trigger \citep{FirstAlert,Bayestar} and we captured our first \DECam{} image of the probability region at 06:00 UTC on August 15, 19.5 hours after the \GW{} detection. \DECam{} is an optical imager, installed on the Blanco 4-m telescope at the Cerro Tololo Inter-American Observatory.  It has a 3-deg$^2$ field of view and is equipped with several broadband optical/NIR filters ({\it u,g,r,i,z,Y,VR}), making it well-suited to search for faint transients over large sky areas \citep{Flaugher}.

\begin{figure*}[htbp!]
\plotone{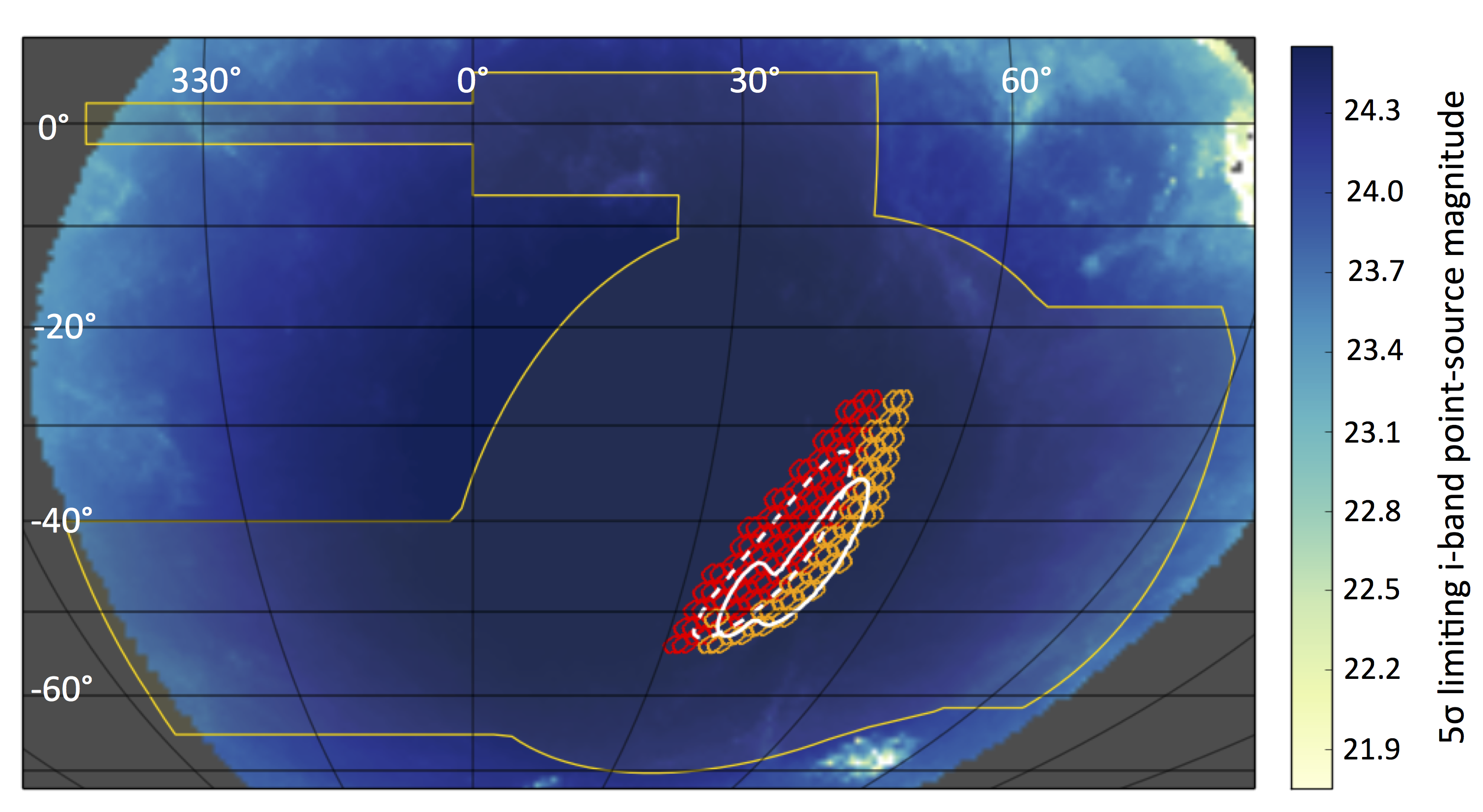}
\caption{Dithered tiling performed for \GW170814 overlaid on the \GW170814 90\%-confidence sky area contours.  The red hexes show the individual pointings that were performed in our search on the first night of observations.  The orange hexes represent the tiles that were not observed until the second night or later due to the sky map change. The white dotted contour shows the initial \texttt{Bayestar} map and the solid white contour represents the final sky map from the LIGO-Virgo O1-O2 GW catalog.  The region enclosed by the yellow contour corresponds to the DES footprint, and the background color shows the estimated 5$\sigma$ point-source limiting magnitude for a 90-second exposure which accounts for air mass and dust extinction \citep[see][]{Neilsen2016}.}
\label{fig:tiles}
\end{figure*}

We imaged the high-probability area of the \Bayestar{} sky map in the {\it i}-band with 90-sec exposures, corresponding to a 5$\sigma$ point-source depth of $\approx 23$ mag. Our strategy of imaging the most probable regions was similar to that used in \cite{SS2016} and \cite{Cowperthwaite2016} (which surveyed in $i$ and $z$ bands), but in order to maximize the sky area coverage we only surveyed in the $i$-band. Our search covered 225 deg$^2$, corresponding to 90\% of the initial \Bayestar{} map, 90\% of the \texttt{LALInference} sky map that was released with the original \GW170814 \LIGO-Virgo publication \citep{GW170814}, and 86\% of the final map from the O1-O2 \GW{} catalog \citep{GWCatalog}.  These estimates account for chip gaps on the camera but not masking of bright stars.  A preliminary \texttt{LALInference} map accounting for calibration uncertainties was sent after our first night of observations, causing a shift in the search region on the second night of observations and onward \citep{PrelimLALInference}. Figure \ref{fig:tiles} shows our tiling over the \LIGO-Virgo sky maps. Observations of the region of interest were taken in epochs which began roughly 0.8, 1.8, 2.8, 5.8, 8.8, 9.8, 10.8, and 11.8 days after the \GW{} event, and each epoch's tiling spanned about 4 hours.  The $5\sigma$ limiting point-source magnitude in the $i$-band was approximately 24 over all the tiles on the first night of observations. The first \DECam{} image was taken on August 15, 2017 at 06:05:31 UTC. Our cadence was chosen to have a dense sampling in time, but observing conditions and follow-up of \GW170817 introduced larger gaps between the third and eighth nights.   

We processed the images from our search using the Dark Energy Survey's (\DES) transient detection pipeline as in \citet{SS2016} and \citet{Herner}. The pipeline consists of a single-epoch processing stage \citep{Morganson,Bernstein} followed by a stage which takes the difference of search images and template images to identify sources with fluctuating brightness \citep[\diffimg,][]{diffimg}.  Template images were available from existing \DES{} data since the \LIGO-Virgo sky maps were contained in the \DES{} footprint.  The sources detected in the pipeline are used to generate candidate light curves: A candidate requires at least two detections by the pipeline, and for each candidate a light curve is constructed from a PSF-fitted flux at each observation. The pipeline also removes persistent point sources in the DES Y1 catalog that are brighter than 20.5 mag in any band.

We split the data into two samples because of a shift in the \GW{} sky map after the first night of observations.  This shift prompted a change in the patch of sky we targeted, creating inhomogeneity in the data sample since the cadence of observations was not uniform over the full area we imaged.  The red and orange ``hexes" in Figure \ref{fig:tiles} show which fields were observed the first night versus only on later nights, respectively. The first dataset, D1, includes the $N_{D1} = 42368$ candidates which were first observed $\sim 0.8$ days after the \GW{} trigger when we were targeting the \Bayestar{} sky map.  D2 contains the $N_{D2}=17192$ candidates observed for the first time after acquiring the preliminary \texttt{LALInference} sky map.  Over the full \GW170814 follow-up campaign, the median number of observations per candidate is 8 and 5 for D1 and D2, respectively.    

%% file: analysis.tex
To identify candidates of interest, we apply selection requirements (or ``cuts") to the full set of candidates produced by \diffimg. We present these criteria in \S\ref{subsec:SR} and have chosen them to (a) minimize contamination from both astrophysical transients such as supernovae and asteroids as well as artifacts in the data and (b) identify ``fast transients'' which quickly decline after the merger. \snana{} simulations \citep{SNANA} of Type Ia and core-collapse SN light curves (using the SALT-II Ia light curve model of \cite{Guy10} and Ibc, IIp, IIn core-collapse templates from \cite{K10}) provide guidance on cuts to remove supernovae. A full optimization and exploration of the cuts is not explored here and is left for future analyses. We choose these cuts using a {\it control sample} of candidates which are away from the highest probability regions of the \texttt{LALInference} sky map, as described in \S\ref{subsec:controlsample}. The number of candidates remaining in the control sample after applying cuts is used to infer the number of candidates expected in the full sample. This inference is detailed in \S\ref{subsec:numpredict}.

Our analysis also makes use of two auxiliary tools: 1) A machine-learning (ML) algorithm, \texttt{autoScan}, trained on \DES{} difference images that produces a score for each difference-imaging detection between 0 and 1 with high scores corresponding to high-confidence point-source-like sources \citep{Goldstein}, and 2) object-fitting algorithms which classify persistent \DES{} sources as galaxies or stars depending on the sources' spatial extent \citep{ADW,Sevilla}.  We use these tools to identify high-confidence point-source detections in our search and to match these detections with stars and galaxies.   

\subsection{Control Sample}\label{subsec:controlsample}
To reduce potential bias in tuning the analysis cuts to reject all events, the cuts are
optimized on a control sample. The control sample comprises a random third of all \diffimg{} candidates, and candidates within 4.5 deg of the maximum a posteriori point of the \citet{GW170814} \texttt{LALInference} sky map are excluded.  There is an $\sim 8\%$ and $\sim 10\%$ chance that the true location of GW170814 is in the control region based on the \citet{GW170814} and \citet{GWCatalog} sky maps, respectively.    

As with the full data set, we split our control sample into two subsamples.  The first subsample C1 (with $N_{C1}=12381$ candidates) comprises the control candidates in D1.  The second subsample C2 contains the control candidates ($N_{C2}=3867$ candidates) in D2.  We apply the cuts in \S\ref{subsec:SR} to the two control subsamples and record the sets of candidates c1 and c2 (with $N_{c1}$ and $N_{c2}$ candidates respectively) passing cuts out of the totals. 

The remaining data (which we call the {\it blinded sample}) is similarly split into two subsamples B1 and B2 for events first observed when targeting the \texttt{Bayestar} map and \texttt{LALInference} map, respectively. In total, Subsample B1 contains $N_{B1}=29987$ candidates and B2 contains $N_{B2} = 13325$ candidates.  Since B1, B2, C1, and C2 are mutually exclusive, we have $N_{D1} = N_{B1}+N_{C1}$ and $N_{D2} = N_{B2}+N_{C2}$. Table \ref{table:ncands} summarizes the numbers of candidates in each subsample.

\begin{table}
\vspace{10pt}
\centering
\begin{tabular}{ll|rr}
\hline
\hline
$N_{D1}$ & 42368 & $N_{D2}$ & 17192\\
$N_{C1}$ & 12381 & $N_{C2}$ & 3867\\
$N_{B1}$ & 29987 & $N_{B2}$ & 13325\\
\hline
\hline
\end{tabular}
\caption{The number of candidates in the two subsets of full (D), control (C), and blinded (B) samples.}
\label{table:ncands}
\end{table}

\subsection{Selection Requirements} \label{subsec:SR}
Below we list the cuts applied to the candidates:

\begin{enumerate}
\item Raw Sample: All candidates produced by \diffimg.
\item 1st Epoch ML$>$0.7: Using the \texttt{autoScan} machine-learning score ($0<{\rm ML}<1$) that was trained with DES data \citep{Goldstein} to remove non-point-source-like detections, we require ${\rm ML}>$ 0.7 for the
first observation. This cut eliminates image artifacts that arise in the difference imaging.  For reference, the \DES{} Supernova program requires ${\rm ML}>0.5$, but for {\it two} separate detections of a candidate rather than just one detection.  Our requirement is more stringent since we are looking for rapidly fading sources and therefore only cut on the first-epoch ML.  Our stricter ${\rm ML}>0.7$ requirement lowers the numbers of single-epoch false positives by a factor of $\sim$ 2 compared with ${\rm ML}>0.5$, while lowering the efficiency by only a few percent at signal-to-noise ratio 10 \citep{Goldstein}.
\item Host Galaxy $z<0.3$: Using galaxies from the \DES{} Y3 Gold catalog, a candidate is matched to a host galaxy if it is within four times the {\it directional light radius} of the galaxy \citep{Gupta}. The directional light radius is the radius of a potential host galaxy in the direction of the candidate transient and is dependent on the survey. Each galaxy is also fit with a Directional Neighborhood Fitting (DNF) photometric redshift $z_{\rm DNF}$ with uncertainty $\Delta z_{\rm DNF}$ \citep{DeVicente}. If the candidate is matched to a galaxy and the best match galaxy satisfies $z_{\rm DNF} - \Delta z_{\rm DNF}>0.3$, the candidate is removed from the sample. This cut removes events that are clearly associated with galaxies beyond the estimated \GW{} redshift of $z= 0.12_{-.04}^{+.03}$. 
\item 2nd Observation S/N $\geq$ 2: The candidate must have a measured signal-to-noise ratio (S/N) of at least 2 on the second observation.  Measurements within one hour of each other are not considered separate observations for this cut. This cut rejects asteroids and difference imaging artifacts.
\item Greater than $2\sigma$ decline: There must be a $>2\sigma$ decline in the flux between the first and second epochs that a candidate was observed. A similar cut was implemented in \citet{SS2016} and \citet{Cowperthwaite2016}. $\sigma$ is the quadrature sum of the flux errors on the two epochs. If multiple measurements of a candidate were taken in the same epoch (i.e. in the same night), we use the first measurement of the epoch. If we did not observe the candidate on the second epoch, it is removed from the sample.  We note that the effect of this cut depends sensitively on the observational choices of the follow-up campaign, not just the astrophysics of the potential \EM{} source.
\item $N_{\rm obs} \geq 4$:  To ensure that we can examine each candidate's light curve over a broad portion of the follow-up campaign, the candidate must have been observed at least $N_{\rm obs} = 4$ times, regardless of S/N.
\item Late-time S/N $< 6$:  After one week from the \GW{} event, the S/N of all observations of a candidate must be less than 6. This requirement removes objects that are bright at late times such as supernovae and variable stars.
\item No Late-time Brightening: To isolate fading transients, we require that after 48 hours from the \GW{} event, there is no increase in flux of the candidate greater than $3\sigma$, where $\sigma$ is the quadrature sum of uncertainties on adjacent flux measurements.
\item Visual Inspection: Subtracted image stamps identified as artifacts (e.g. cosmic rays) are removed from the sample.
\end{enumerate}
After applying these cuts to the control sample, $N_{c1}=1$ and $N_{c2}=0$ candidates remain.

\subsection{Expectation of Number of Candidates in Full Sample}\label{subsec:numpredict}
Given $N_{c1}$ and $N_{c2}$ out of $N_{C1}$ and $N_{C2}$ candidates passing in the control fields, respectively, we expect $\langle N_{b1}+N_{b2}\rangle = N_{c1}N_{B1}/N_{C1} + N_{c2}N_{B2}/N_{C2} = 2.4$ events in B1+B2, which we interpret as the mean of a Poisson distribution. This interpretation does not account for small differences in Milky Way reddening and stellar density over the search region. In \S\ref{sec:results}, we analyze the blinded sample and compare our expectations to the number of candidates passing the cuts.

%% file: results.tex
\begin{figure*}
\epsscale{1.1}
\centering
\plotone{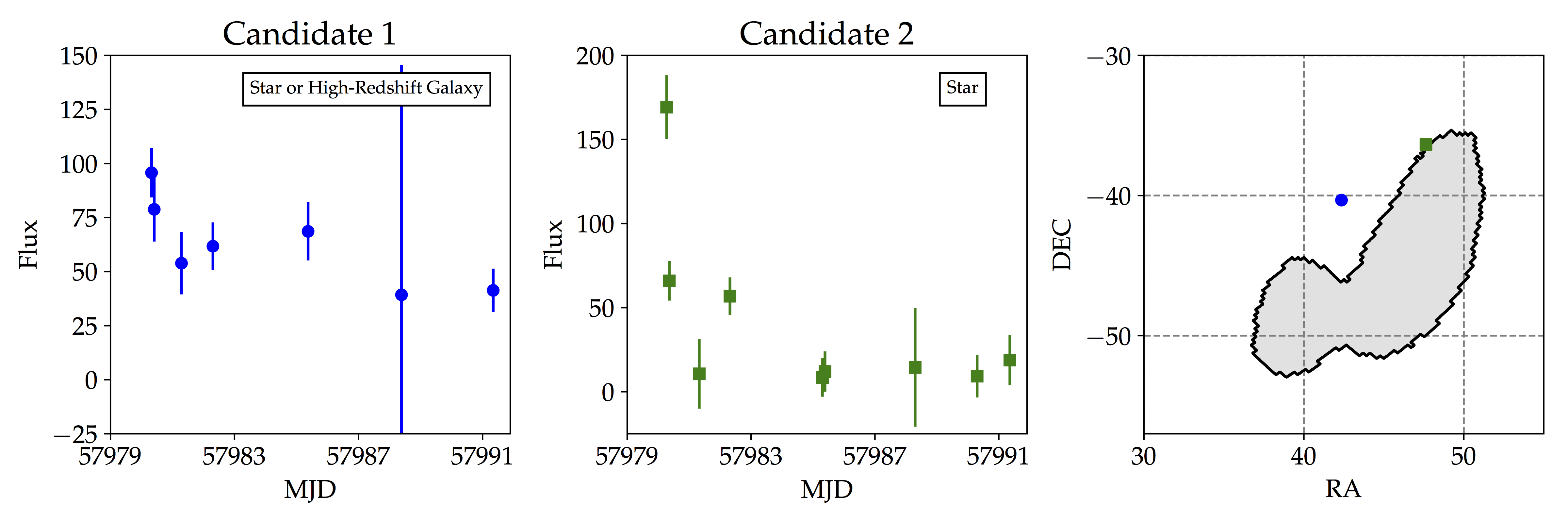}
\epsscale{1.1}
\plotone{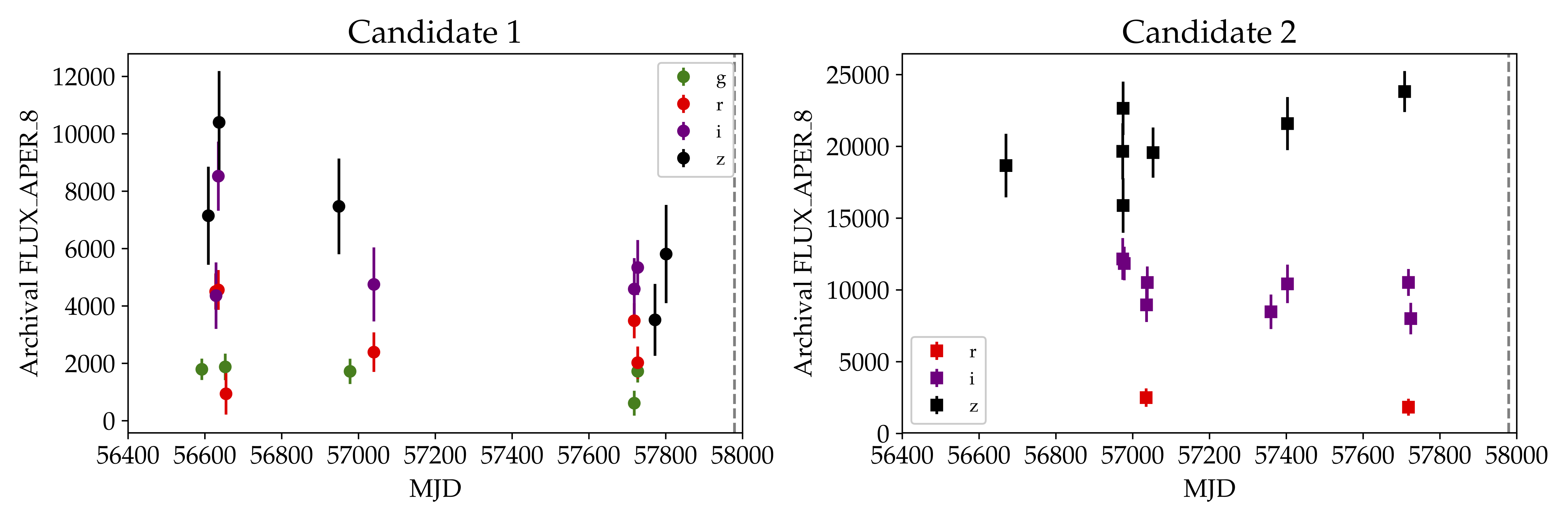}
\caption{Light curves, archival fluxes, and sky positions for the two candidates passing all cuts.  {\it Top:} The left panel shows the $i$-band light curve for Candidate 1 (associated with Y3 Gold star or high-redshift galaxy), and the middle panel shows the same for Candidate 2 (associated with Y3 Gold star).  The flux is defined in relation to AB magnitude as $m_{\rm AB} = -2.5\log_{10}({\rm Flux}) + 27.5$.  The right panel overlays the sky positions of the two candidates on the 90\% credible region of the \texttt{lalinference} sky map (gray). {\it Bottom:} Available archival flux measurements in {\it g,r,i,z} bands at the locations of Candidates 1 (left) and 2 (right).  These \texttt{FLUX\_APER\_8} fluxes are taken with 22.22-pixel apertures and are not from difference imaging and therefore cannot be directly compared to those in the top panels.  The vertical, gray, dashed line on the far right of the two plots indicates the GW170814 merger time.}
\label{fig:light_curves}
\end{figure*}

Table \ref{table:cuts} shows the effect of the cuts on the full sample, which includes the control sample.  It also shows the initial $i$-band magnitudes and sky positions for the events passing all cuts.  After analyzing the blinded sample, one more candidate is found, leaving a total of two candidates passing cuts in the control and blinded samples, with ID numbers \IDcontrol{} and \IDblind{}, respectively. Finding one candidate passing cuts in the blinded sample is consistent with the 2.4 expected background events derived from the control sample presented in \S\ref{subsec:numpredict}. The light curves for both events and their sky positions are shown in Figure \ref{fig:light_curves}.  

\begin{figure}
\epsscale{1.2}
\plotone{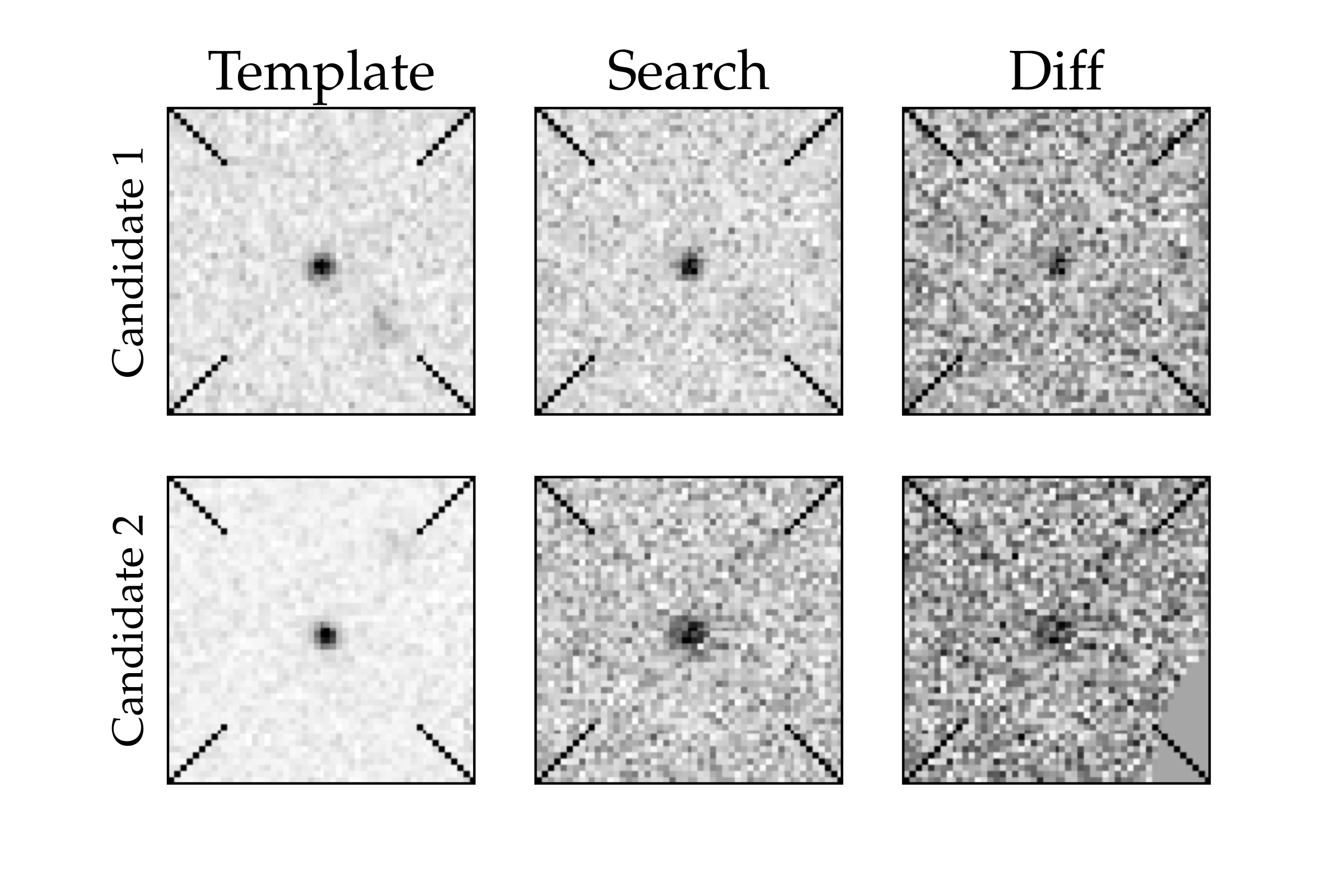}
\caption{Template, search, and difference image stamps for candidates passing cuts.  The top row shows the $i$-band images for Candidate \IDcontrol{}, and the bottom for Candidate \IDblind.  The search and difference images are from the the first epoch of observations of the candidate.  Each stamp is 13.2" x 13.2".}
\label{fig:stamps}
\end{figure}

Upon visual inspection of the two candidates, neither is an obvious subtraction artifact or cosmic ray. Here we do not show examples of subtraction artifacts and cosmic rays that would be cut by visual inspection since visual inspection did not end up removing any candidates in this analysis.  However, the template images for both candidates contain a bright source at the position of the candidates.  The template, search and difference images from the first epoch of observations of each candidate are shown in Figure \ref{fig:stamps}.

A deeper search through the DES high-quality object catalog (``Y3 Gold") reveals that Candidate \IDcontrol{} is associated with an object that is classified as either a galaxy at $z\sim0.9$, or a star, depending on the classifier used. A multi-epoch, multi-object fitting algorithm classifies the object's PSF as a candidate star, whereas the single-object fit categorizes the object as a galaxy \citep{ADW,Sevilla}. Notably, the object is too faint to meet the brightness cutoff for inclusion in our star veto catalog and it is not vetoed by our host galaxy redshift cut (cut 3) because we only include the highest confidence galaxies in the host-galaxy matching.  Fitting each band to a constant flux for all archival observations of the object (\DES{} Years 1-4) results in a $\chi^2$/DOF of $48.6/17 = 2.9$ and $p$-value $p(\chi^2\geq 48.6|{\rm DOF}=17)=7\times10^{-5}$, indicating previous variability of the source.  These archival fluxes are shown in Figure \ref{fig:light_curves}. Spectroscopic observations of this source could clarify if the object is a star or galaxy. 

Candidate \IDblind{} is also associated with a \DES{} Y3-Gold object and is classified as a star by both classifiers and constant-flux fits to archival observations yield a $\chi^2$/DOF of $25.7/14=1.8$ and $p$-value of $p(\chi^2\geq 25.7|{\rm DOF}=14)=0.03$ (see Figure \ref{fig:light_curves}). However, the star is also too faint (by 0.16 mag) to meet the brightness cutoff for the star veto catalog of our pipeline and hence was not removed by the 20.5 mag persistent-point-source cut in \S\ref{sec:search}.



\begin{table}
\begin{tabular}{llll}
Cuts & $N_{\rm seq}$\footnote{Number of candidates remaining after applying each cut sequentially} & $N_{\rm only}\footnote{Number of candidates after applying an individual cut}$ & $N_{\rm LO}\footnote{Number of candidates if a cut is ``left out'' but all the rest are applied}$ \\ \hline
1. Raw Sample & 59560 & -- & -- \\ \hline
2. 1st Epoch ML $>$ 0.7 & 1206 & 1206 & 258 \\
3. Unmatched or Host  $z<$ 0.30 & 730 & 31119 & 8 \\
4. 2nd Obs S/N $\geq$ 2.0 & 663 & 44181 & 4 \\
5. $>2.0$ sigma decline & 45 & 5570 & 65 \\
6. $N_{\rm obs}\geq4$ & 31 & 50029 & 2 \\
7. Late-time S/N $<$ 6 & 4 & 27571 & 21 \\
8. No Late-time Brightening & 2 & 36499 & 4 \\
9. Visual Inspection & 2 & -- & 2 \\ \hline \vspace{0.5cm}
\end{tabular}
\begin{tabular}{llll}
Candidate \# & RA & DEC & $m_i$\\ \hline
\IDcontrol & 42.35047$^\circ$ & -40.32632$^\circ$ & 22.5\\ \hline
\IDblind & 47.63365$^\circ$ & -36.36045$^\circ$ & 21.9\\ \hline
\end{tabular}
\caption{{\it Top}: Candidates remaining in the full data sample after applying cuts.  {\it Bottom}: sky coordinates and initial $i$-band magnitude $m_i$ of the two candidates passing all cuts.}
\label{table:cuts}
\end{table}

%% file: discussion.tex
Although our search identified two interesting candidates, it is unlikely that either candidate is associated with \GW170814.  Neither candidate is located in the 90\% confidence region of the \texttt{LALInference} sky map, and both are associated with existing objects in \DES{} catalogs that are inconsistent with our expectations of the \GW{} source. Candidate \IDblind{} is likely the transient behavior of a variable star and is consistent with the number of background candidates expected in the blinded sample.  Candidate \IDcontrol{} could also be stellar variability, or it could be a signal associated with a distant galaxy.  Assuming it is a galaxy, the DNF photometric redshift of the object is $z=0.95\pm0.12$, far beyond the possible redshift of \GW170814 at that sky position: The 99\% upper limit on the \GW{} distance along the Candidate \IDcontrol{} line of sight is 615 Mpc, whereas the galaxy distance is $6380_{-980}^{+1010}$ Mpc assuming the LCDM cosmology parameters of \cite{Planck15}. We note however that photometric redshifts can occasionally have catastrophic failures. 

An alternative explanation for the persistent emission from the two candidates is that one or both of these candidates is associated with a quasar.  If either is a quasar, it is unlikely to be at the low redshifts of interest for GW170814 \citep{Paris}. Spectroscopic follow-up of the persistent sources associated with Candidates 1 and 2 could resolve whether we have mis-categorized them.

We conclude that these two candidates are not associated with \GW170814, and thus we find no \EM{} counterpart associated with the \BBH{} merger over the \area{} deg$^2$ region that we surveyed with 86\% sky map coverage.  We have not yet computed the efficiency, which is needed to set rate limits on BBH merger emission, but this rate-limit analysis is underway using \texttt{SNANA} simulations similar to those used in \cite{SS2016}.  Our rate-limit analysis will also re-evaluate the cuts to maximize possible \BBH{} model efficiency while minimizing supernova background events.  Qualitatively though, the analysis presented here covers 86\% of the \GW{} sky map and searches for events with rapidly declining light curves. The non-detection of an \EM{} counterpart in our sample results in stringent limits on fast-declining optical models brighter than $i \sim 23$ mag 1-2 days after the \BBH{} coalescence. This search is not sensitive to models that fade faster than the time between the first two observations due to Cut 4. Assuming a flat-in-frequency optical spectrum from 4000 \AA{} to 7000 \AA{} and the \GW170814 median distance, this $i \sim 23$ mag limit corresponds to a luminosity limit of $L_{\rm optical} \sim 5\times10^{41}$ erg/s.  

Our results constrain the space of models put forth in e.g.~\citet{Stone}, \citet{deMink}. For example, \citet{Stone} posits that \BBH{} mergers occurring in the gaseous environments of {\small AGN} disks could be accompanied by gas accretion onto the final merged black hole that powers luminosities of order $L\sim 10^{40}$ erg/s lasting a few years, but highly super-Eddington accretion might result in a brighter and shorter-lived transient that our analysis is sensitive to. Our search also narrows the feasibility of models from \citet{deMink}, which predict emission with luminosities of approximately $L\sim 10^{42}$ erg/s occurring on fast timescales. The search performed here is tailored to remove longer-lived transients, and therefore it does not constrain long-lived \BBH{} counterparts, such as the supernova association suggested in \citet{Loeb} and later discussed in \citealt{Woosley} and \citealt{DOrazio}.


Aside from identifying interesting candidates, our search for counterparts to \GW170814 is a test-bed for future BBH follow-up analyses where the sky map credible areas will be small enough to be completely tiled in less than one night using \DECam{}.  For a real-time search for future counterparts, we consider resources to spectroscopically follow $\sim 10$ candidates, which we would want to identify within roughly two days of the \GW{} trigger.  In this scenario, we only apply the first five cuts in Table \ref{table:cuts}, since the remaining cuts depend on observations beyond two days.  Through cut 5 ($>2\sigma$ decline), our search finds 45 candidates.  Of these, we find that four candidates (including Candidates 1 and 2) are associated with \DES{}-catalog objects that are either galaxies beyond our redshift cut (cut 3), or stars, and are thus uninteresting as black-hole-merger counterparts. Excluding these four candidates, our real time search would find 41 candidates over \area{} deg$^2$ or $\sim 16$ candidates per 87 deg$^2$ (the 90\% credible area of the final \GW170814 sky map). For comparison, \citet{Cowperthwaite15} predicts $\sim 19$ Type Ia SNe detected at $z<0.25$ over a 7-day, 87-deg$^2$ search. This suggests that the first five cuts are adequate to find interesting spectroscopic targets over a region the size of the \GW170814 sky map. 

Future work will incorporate simulations of \BBH{} and \SN{} light curves to assess the efficiency and false alarm rate of our search. If several \BBH{} events are followed up with no \EM{} counterpart found, a combined analysis will be needed to set limits on \BBH{} \EM{} emission.

%% file: conclusion.tex
We have presented an optical search for counterparts to gravitational wave \GW170814 using the Dark Energy Camera. Our search covered \area{} deg$^2$, corresponding to 86\% of the final \LIGO-Virgo sky map.  Our difference-imaging pipeline produces 59560 light curves from the search images which are analyzed with the criteria in \S\ref{subsec:SR}.  After applying these cuts to the {\it i}-band light curves, two candidates remain.  These two candidates are most likely not associated with \GW170814: one is a high-confidence variable star, and the other is either a variable star or a transient associated with a high-redshift galaxy well beyond the expected \GW170814 redshift.   

With no candidates associated with \GW170814, our analysis disfavors fast-declining optical emission from \BBH{} mergers 1 to 2 days after merger with $i \lesssim 23$ mag.  Future work will assess the efficiency and false-positive rate in optical \BBH{} searches such as this one using simulations of \BBH{} and \SN{} light curves.  Additionally, we will consider updates to our star veto catalog and galaxy catalog to account for fainter stars and objects with uncertain star or galaxy classification.  

Tens of \BBH{} signals are expected in the \LVC's third operating run, and some are likely to have localization regions of similar size to that of \GW170814.  Based on the search and analysis presented here, we are preparing to search for additional \BBH{} merger signals and quickly identify candidates for spectroscopic follow up.  With future \BBH{} optical searches and forward modeling of background and foreground signals, we will set increasingly stringent limits on \BBH{} \EM{} emission.  Although  \BBH{} mergers may remain electromagnetically dark, the future of \BBH{} astrophysics is bright.   

%% file: ack.tex
ZD would like to thank Reed Essick for useful discussions and would also like to thank the Dark Cosmology Center and Neils Bohr Institute where he was studying at the time of the GW170814 alert.  ZD is supported by NSF Graduate
Research Fellowship grant DGE-1144082. ZD and DEH were partially supported by NSF CAREER grant PHY1151836
and NSF grant PHY-1708081. H.-Y.C. was supported by the Black Hole Initiative at Harvard University, through a grant from the John Templeton Foundation. This work was completed in part with resources provided by the University of Chicago Research Computing Center and support from the Kavli Institute for Cosmological Physics at the University of Chicago through NSF grant PHY-1125897 and an endowment from the Kavli Foundation. 

Funding for the DES Projects has been provided by the DOE and NSF (USA), MEC/MICINN/MINECO (Spain), STFC (UK), HEFCE (UK). NCSA (UIUC), KICP (U. Chicago), CCAPP (Ohio State), 
MIFPA (Texas A\&M), CNPQ, FAPERJ, FINEP (Brazil), DFG (Germany) and the Collaborating Institutions in the Dark Energy Survey.

The Collaborating Institutions are Argonne Lab, UC Santa Cruz, University of Cambridge, CIEMAT-Madrid, University of Chicago, University College London, 
DES-Brazil Consortium, University of Edinburgh, ETH Z{\"u}rich, Fermilab, University of Illinois, ICE (IEEC-CSIC), IFAE Barcelona, Lawrence Berkeley Lab, 
LMU M{\"u}nchen and the associated Excellence Cluster Universe, University of Michigan, NOAO, University of Nottingham, Ohio State University, University of 
Pennsylvania, University of Portsmouth, SLAC National Lab, Stanford University, University of Sussex, Texas A\&M University, and the OzDES Membership Consortium.

Based in part on observations at Cerro Tololo Inter-American Observatory, National Optical Astronomy Observatory, which is operated by the Association of 
Universities for Research in Astronomy (AURA) under a cooperative agreement with the National Science Foundation.

The DES Data Management System is supported by the NSF under Grant Numbers AST-1138766 and AST-1536171. 
The DES participants from Spanish institutions are partially supported by MINECO under grants AYA2015-71825, ESP2015-66861, FPA2015-68048, SEV-2016-0588, SEV-2016-0597, and MDM-2015-0509, 
some of which include ERDF funds from the European Union. IFAE is partially funded by the CERCA program of the Generalitat de Catalunya.
Research leading to these results has received funding from the European Research
Council under the European Union's Seventh Framework Program (FP7/2007-2013) including ERC grant agreements 240672, 291329, and 306478.
We  acknowledge support from the Australian Research Council Centre of Excellence for All-sky Astrophysics (CAASTRO), through project number CE110001020, and the Brazilian Instituto Nacional de Ci\^encia
e Tecnologia (INCT) e-Universe (CNPq grant 465376/2014-2).

This manuscript has been authored by Fermi Research Alliance, LLC under Contract No. DE-AC02-07CH11359 with the U.S. Department of Energy, Office of Science, Office of High Energy Physics. The United States Government retains and the publisher, by accepting the article for publication, acknowledges that the United States Government retains a non-exclusive, paid-up, irrevocable, world-wide license to publish or reproduce the published form of this manuscript, or allow others to do so, for United States Government purposes.